\documentclass[entropy,article,submit,moreauthors,pdftex,10pt,a4paper]{mdpi}
\firstpage{1}
\articlenumber{x}
\doinum{10.3390/------}
\pubvolume{xx}
\pubyear{2017}
\copyrightyear{2017}
\externaleditor{Academic Editor: name}
\history{Received: date; Accepted: date; Published: date}

\pdfoutput=1

\usepackage[T1]{fontenc}
\usepackage[utf8]{inputenc}
\usepackage[main=english,italian]{babel}

\usepackage{amssymb,dsfont,amscd,mathrsfs}
\usepackage[subnum]{cases}
\usepackage[centercolon=true]{mathtools}
\usepackage{braket,slashed,bookmark,xspace}

\usepackage{csquotes}
\usepackage[capitalize]{cleveref}
\hypersetup{
  linkcolor=red,
  colorlinks=true
}

\newcommand*{\swap}[2]{\let\temp#1 \let#1#2 \let#2\temp \let\temp\relax}
\swap{\epsilon}{\varepsilon}
\swap{\theta}{\vartheta}
\swap{\phi}{\varphi}

\DeclareMathOperator{\Arccos}{Arccos}

\newcommand*{\ketbra}[2]{\ket{#1}\!\bra{#2}}

\newcommand*{\mvec}[1]{\begin{pmatrix}#1\end{pmatrix}}

\newcommand*{\deq}{\coloneqq}
\newcommand*{\eqd}{\eqqcolon}
\newcommand*{\dd}{\mathrm{d}}

\newcommand*{\ct}[1]{{#1}^{\dagger}}
\newcommand*{\cg}[1]{
   \vbox{%
     \hrule height 0.5pt
     \kern0.25ex
     \hbox{%
       \kern-0.1em
       \ifmmode#1\else\ensuremath{#1}\fi
       \kern-0.1em
     }
   }
}

\newcommand*{\UP}{{\uparrow}}
\newcommand*{\DOWN}{{\downarrow}}

\newcommand*{\vv}[1]{{\boldsymbol{#1}}}

\newcommand*{\bv}{\vv{v}}
\newcommand*{\bw}{\vv{w}}

\newcommand*{\mbb}[1]{{\mathbb{#1}}}

\newcommand*{\Z}{\mbb{Z}}
\newcommand*{\R}{\mbb{R}}
\renewcommand*{\C}{\mbb{C}}

\newcommand*{\sns}[1]{{\mathsf{#1}}}
\newcommand*{\Br}{\sns{B}}
\newcommand*{\sS}{\sns{S}}

\newcommand*{\scr}[1]{\mathscr{#1}}
\renewcommand*{\H}{\scr{H}}

\newcommand*{\eg}{\emph{e.g.}\xspace}
\newcommand*{\ie}{\emph{i.e.}\xspace}

\newcommand*{\tp}[1]{_{\textup{#1}}}

\Title{Solutions of a two-particle interacting quantum walk}

\orcidauthorONE{0000-0000-000-000X} 

\Author{Alessandro Bisio $^{1}$, Giacomo
  Mauro D'Ariano $^{1}$, Nicola Mosco $^{1}$*, Paolo
  Perinotti $^{1}$*, Alessandro Tosini $^{1}$}

\AuthorNames{Alessandro Bisio, Giacomo Mauro D'Ariano, Nicola Mosco, Paolo Perinotti and Alessandro Tosini}

\address{%
$^{\dag}$ \quad Dipartimento di Fisica dell’Universit\`a di Pavia, via Bassi 6, 27100 Pavia and
Istituto Nazionale di Fisica Nucleare, Gruppo IV, via Bassi 6, 27100
Pavia}

\corres{Correspondence: paolo.perinotti@unipv.it}

\abstract{We study the solutions of the interacting Fermionic cellular automaton
introduced in Ref. \cite{PhysRevA.97.032132}. The automaton is the analogue of
the Thirring model with both space and time discrete. We present a  derivation
of the two-particles solutions of the automaton, which exploits the symmetries
of the evolution operator. In the two-particles sector, the evolution operator
is given by the sequence of two steps, the first one corresponding to a unitary
interaction activated by two-particle excitation at the same site, and the
second one to  two independent one-dimensional Dirac quantum walks. The
interaction step can be regarded as the discrete-time version of the interacting
term of some Hamiltonian integrable system, such as the Hubbard or the Thirring
model. The present automaton exhibits scattering solutions with nontrivial
momentum transfer, jumping between different regions of the Brillouin zone that
can be interpreted as Fermion-doubled particles, in stark contrast with the
customary momentum-exchange of the one dimensional Hamiltonian systems. A
further difference compared to the Hamiltonian model is that there exist bound
states for every value of the total momentum, and even for vanishing coupling
constant. As a complement to the  analytical derivations we show numerical
simulations of the interacting evolution.}



\begin{document}

\section{Introduction}

Quantum walks (QWs) describe the evolution of one-particle quantum states on a
lattice, or more generally, on a graph. The quantum walk evolution is linear in the
quantum state and the quantum aspect of the evolution occurs in the interference
between the different paths available to the walker. There are two kinds of
quantum walks: continuous time QWs, where the evolution operator of the system
given in terms of an Hamiltonian can be applied at any time (see Farhi et
al.~\cite{Farhi:2008aa}), and discrete-time QWs, where the evolution operator is
applied in discrete unitary time-steps. The discrete-time model, which appeared already
in the Feynman discretization of the Dirac equation~\cite{Feynman:1965aa}, was
later rediscovered in quantum
information \cite{grossing1988quantum,AB01,reitzner2011quantum,gross2012index,Shikano:2013:1546-1955:1558},
and proved to be a versatile platform for various scopes. For example, QWs have
been used for empowering quantum algorithms, such as database search
\cite{PhysRevA.70.022314,portugal2013quantum}, or graph isomorphism
\cite{1751-8121-41-7-075303,PhysRevA.81.052313}.  Moreover, quantum walks have
been studied as a simulation tool for relativistic quantum fields
\cite{bialynicki1994weyl,meyer1996quantum,Yepez:2006p4406,Arrighi:2013ab,Bisio:2015aa,PhysRevA.90.062106,DAriano:2014ad,DAriano:2015aa,Arrighi2016,Bisio:2016aa,PhysRevA.93.052301,Bisio:2016ac,Mallick2017,1367-2630-18-10-103038,mlodin1,mlodin2},
and they have been used as discrete models of spacetime
\cite{BBDPT15,Bisio:2016ab,Bisio20150232,1367-2630-16-9-093007}.

QWs are among the most promising quantum simulators with possible realizations
in a variety of physical systems, such as nuclear magnetic resonance
\cite{PhysRevA.67.042316,PhysRevA.72.062317}, trapped
ions \cite{PhysRevLett.103.183602}, integrated photonics, and bulk optics
\cite{Do:05,PhysRevLett.108.010502,Crespi:2013aa,flamini2018photonic}.

New research perspectives are unfolding in the scenario of multi-particle
interacting quantum walks where two or more walking particles are coupled via
non-linear (in the field) unitary operators. The properties of these systems are
still largely unexplored. Both continuous-time~\cite{Childs:2009aa} and
discrete-time~\cite{PhysRevA.81.042330} quantum walks on sparse unweighted
graphs are equivalent in power to the quantum circuit model. However, it is
highly non trivial to design a suitable architecture for universal quantum
computation based on quantum walks. Within this perspective a possible route has
been suggested in Ref.~\cite{childs2013universal} based on interacting
multi-particle quantum walks with indistinguishable particles (Bosons or
Fermions), proving that ``almost any interaction'' is universal. Among the
universal interacting many-body systems are the models with coupling term of the
form $\chi \delta_{x_1,x_2} \hat{n}(x_1)\hat{n}(x_2)$, with $\hat{n}(x)$ the
number operator at site $x$.  The latter two-body interaction lies at the basis of
notable integrable quantum systems in one space dimension such as the Hubbard and the
Thirring Hamiltonian models.

The first attempt at the analysis of interacting quantum walks was carried out in
Ref.~\cite{1367-2630-14-7-073050}. More recently, in
Ref.~\cite{PhysRevA.97.032132}, the authors proposed a discrete-time analogue of
the Thirring model which is indeed a Fermionic quantum cellular automaton, whose dynamics in the two-particles sector reduces to an interacting two-particle quantum walk. As for its Hamiltonian counterpart, the
discrete-time interacting walk has been solved analytically in the case of two
Fermions.  Analogously to any Hamiltonian integrable system, also in the
discrete-time case the solution is based on the Bethe Ansatz technique.
However, discreteness of the evolution prevents the application of the usual
Ansatz, and a new Ansatz has been introduced successfully~\cite{PhysRevA.97.032132}.

In this paper we
present an original simplified derivation of the solution of
Ref.~\cite{PhysRevA.97.032132} which exploits the symmetries of the interacting
walk.  We present the diagonalization of the evolution operator and the
characterization of its spectrum. We explicitly write the two particle
states corresponding to the scattering solutions of the system, having
eigenvalues in the continuous spectrum of the evolution operator.  We then
show how the present model predicts the formation of bound states, which are
eigenstates of the interacting walk corresponding to the discrete spectrum.  We
provide also in this case the analytic expression of such molecular states.

We remark the phenomenological differences between the Hamiltonian model and the
discrete-time one. First we see that the set of possible scattering solutions is
larger in the discrete-time case: for a fixed value total momentum, a non
trivial transfer of relative momentum can occur besides the simple exchange of
momentum between the two particles, differently from the Hamiltonian case. Also
the family of bound states appearing in the discrete-time scenario is larger
than the corresponding Hamiltonian one.  Indeed, for any fixed value of the
coupling constant, a bound state exists with any possible value of the total
momentum, while for Hamiltonian systems bound states cannot have arbitrary total
momentum.

Finally we show that in the set of solutions for the interacting walk there
are perfectly localized states (namely states which lie on a finite number of
lattice sites) and, differently from the Hamiltonian systems, bound states exist also
for vanishing coupling constant. In addition to the exact
analytical solution of the dynamics we show the simulation of some significant
initial state.

\section{The Dirac Quantum Walk}

In this section, we review the Dirac walk on the line describing the free
evolution of a two-component Fermionic field. The walk evolution is provided by
the unitary operator $W$ acting on the single particle Hilbert space $\H\deq\C^2
\otimes \ell^2(\Z)$ for which we employ the factorized basis $\ket{a}\ket{x}$,
with $a\in\{\UP,\DOWN\}$ and $x\in\Z$. Being the evolution of a quantum walk
linear in the field operators, the single-step evolution is expressed by the
following equation:
\begin{align*}
  \psi(x,t+1) = W \psi(x,t), \qquad
    \psi(x,t) = \mvec{\psi_\UP(x,t)\\\psi_\DOWN(x,t)}
\end{align*}
with $W$ given by
\begin{equation} \label{eq:dirac-1d-op}
  W =
  \begin{pmatrix}
    \nu\ct T_x & -i\mu \\
    -i\mu      & \nu T_x
  \end{pmatrix}, \qquad
  \nu,\mu > 0, \quad \nu^2 + \mu^2 = 1,
\end{equation}
where $T_x$ denotes the translation operator on $\ell^2(\Z)$, defined by
$T_x\ket{x} = \ket{x+1}$.

Since the walk $W$ is translation invariant (it commutes with the translation
operator), it can be diagonalized in momentum space. In the momentum
representation, defining $\ket p \deq (2\pi)^{-1/2}\sum_{x\in\Z} e^{-ipx} \ket
x$, with $p\in\Br\deq(-\pi,\pi]$, the walk operator can be written as
\begin{align*}
  W = \int_\Br \!\!\dd p \; W(p) \otimes \ketbra{p}{p}, \qquad
  W(p) =
  \begin{pmatrix}
    \nu e^{ip} & -i\mu \\
    -i\mu & \nu e^{-ip}
  \end{pmatrix},
\end{align*}
where $|\nu|^2+|\mu|^2=1$.
The spectrum of the walk is given by $\{e^{-i\omega(p)},\, e^{i\omega(p)}\}$,
where the dispersion relation $\omega(p)$ is given by
\begin{equation} \label{eq:dirac-1d-dispe}
  \omega(p) \deq \Arccos(\nu\cos p),
\end{equation}
where $\Arccos$ denotes the principal value of the arccosine function.
The single-particle eigenstates, solving the eigenvalue problem
\begin{gather} \label{eq:dirac-1d-eigen}
  W(p) \bv^s_p = e^{-is\omega(p)} \bv^s_p\,, \quad s=\pm,
\end{gather}
can be conveniently written as
\begin{gather*}
  \bv^s_p = \frac{1}{|N_s|}
    \mvec{-i\mu\\g_s(p)},
\end{gather*}
with $g_s(p) \deq -i(s\sin\omega(p) + \nu\sin p)$, $|N_s|^2 \deq \mu^2 +
|g_s|^2$.

\section{The Thirring Quantum Walk} \label{sec:thir}

In this section we present a Fermionic cellular automaton in one spatial
dimension with an on-site interaction, namely two particles interact only when
they lie at the same lattice site. The linear part corresponds to the Dirac
QW~\cite{DAriano:2014ae} and the interaction term is the most general
number-preserving coupling in one dimension~\cite{Ostlund:1991aa}. The same kind
of interaction characterizes also the most studied integrable quantum systems,
such as the Thirring~\cite{Thirring:1958aa} and the
Hubbard~\cite{Hubbard:1963aa} models.

The linear part of the $N$-particle walk is described by the operator $W_N \deq
W^{\otimes N}$, acting on the Hilbert space $\H_N = \H^{\otimes N}$ and
describing the free evolution of the particles. In order to introduce an
interaction, we modify the update rule of the walk with an extra step
$V\tp{int}$: $U_N \deq W_N V\tp{int}$. In the present case the term $V\tp{int}$
has the form
\begin{equation*}
  V\tp{int} = V_N(\chi): = e^{i\chi n_\UP(x) n_\DOWN(x)},
\end{equation*}
where $n_a(x)$, $a \in \{\UP,\DOWN\}$, represents the particle number at site
$x$, namely $n_a(x) = \ct\psi_a(x)\psi_a(x)$, and $\chi$ is a real coupling
constant. Since the interaction term preserves the total number operator we can
study the walk dynamics for a fixed number of particles. In this work we focus
on the two-particle sector whose solutions has been derived in
Ref.~\cite{PhysRevA.97.032132}. As we will see, the Thirring walk features molecule
states besides scattering solutions. This features is shared also by the
Hadamard walk with the same on-site interaction~\cite{Ahlbrecht:2011ab}.

Since we focus on the solutions involving the interaction of two
particles, it is convenient to write the walk in the centre of mass
basis $\ket{a_1,a_2}\ket{y}\ket{w}$, with
$a_1,\, a_2 \in \{\UP,\DOWN\}$, $y = x_1 - x_2$ and $w = x_1 + x_2$.
Therefore in this basis the generic Fermionic state is
$\ket{\psi}=\sum_{a_1,a_2,y,w}c(a_1,a_2,y,w)\ket{a_1,a_2}\ket{y}\ket{w}$
with $c(a_2,a_1,y,w)=-c(a_1,a_2,-y,w)$.  Notice that only the pairs
$y,w$ with $y$ and $w$ both even or odd correspond to physical points
in the original basis $x_1,x_2$.

We define the two-particle walk with both $y$ and $w$ in $\Z$, so that
the linear part of walk can be written as
\begin{gather} \label{eq:th-centre-mass}
  W_2 = \mu\nu
  \begin{pmatrix}
    \frac{\nu}{\mu} T_w^2 &
    -i T_y \otimes T_w &
    -i \ct T_y \otimes T_w &
    -\frac{\mu}{\nu} \\
    -i T_y \otimes T_w &
    \frac{\nu}{\mu} T_y^2 &
    -\frac{\mu}{\nu} &
    -i \ct T_y \otimes T_w \\
    -i \ct T_y \otimes T_w &
    -\frac{\mu}{\nu} &
    \frac{\nu}{\mu} {\ct T_y}^2 &
    -i \ct T_y \otimes \ct T_w \\
    -\frac{\mu}{\nu} &
    -i T_y \otimes \ct T_w &
    -i \ct T_y \otimes \ct T_w &
    \frac{\nu}{\mu} {\ct T_w}^2
  \end{pmatrix},
\end{gather}
where $T_y$ represents the translation operator in the relative
coordinate $y$, and $T_w$ the translation operator in the centre of
mass coordinate $w$, whereas the interacting term reads
\begin{equation*}
  V_2(\chi) =
  \begin{pmatrix}
    I_y \otimes I_w & 0 & 0 & 0 \\
    0 & e^{i\chi\delta_{y,0}} \otimes I_w & 0 & 0 \\
    0 & 0 & e^{i\chi\delta_{y,0}} \otimes I_w & 0 \\
    0 & 0 & 0 & I_y \otimes I_w
  \end{pmatrix}.
\end{equation*}
This definition gives a walk $U_2=W_2V_2(\chi)$ that can be decomposed
in two identical copies of the original walk. Indeed, defining as
$C$ 
the projector on the physical center of mass coordinates, one has
$U_2=CU_2C+(I-C)U_2(I-C)$,
where $CU_2C$
and $(I-C)U_2(I-C)$
are unitarily equivalent. We will then diagonalize the operator
$U_2$,
reminding that the physical solutions will be given by projecting the
eigenvectors with $C$.
 
Introducing the (half) relative momentum $k = \frac{1}{2}(p_1 - p_2)$
and the (half) total momentum $p = \frac{1}{2}(p_1 + p_2)$, the free
evolution of the two particles is written in the momentum
representation as
\begin{align*}
  & W_2 = \int\!\! \dd k \dd p \; W_2(p,k) \otimes \ketbra{k}{k} \otimes
    \ketbra{p}{p}, \\
\intertext{where the matrix $W_2(p,k)$ is defined as}
  & W_2(p,k) \deq W(p+k) \otimes W(p-k). \\
\intertext{Furthermore, we introduce the vectors $\bv_k^{sr} \deq \bv_{p+k}^s
\otimes \bv_{p-k}^r$, with $s,\, r = \pm$, such that}
  & W_2(p,k) \bv_k^{sr} = e^{-i\omega_{sr}(p,k)} \bv_k^{sr},
\end{align*}
where $\omega_{sr}(p,k) \deq s\omega(p+k) + r\omega(p-k)$ is the dispersion
relation of the two-particle walk. Explicitly, the vectors $\bv_k^{sr}$ are
given by
\begin{equation} \label{eq:dirac-eigen-free}
  \bv_k^{sr} =
  \frac{1}{|N_s(p+k)| \, |N_r(p-k)|}
  \mvec{
    -\mu^2 \\
    -i\mu g_r(p-k) \\
    -i\mu g_s(p+k) \\
    g_s(p+k)g_r(p-k)
  }.
\end{equation}

We focus in this work on Fermionic solutions satisfying the eigenvalue equation
\begin{equation} \label{eq:interac-eigen}
  U_2(\chi,p) \ket\psi = e^{-i\omega} \ket\psi, \qquad \omega \in \R,
\end{equation}
with $\ket{\psi(y)} \in \C^4$. In the centre of mass basis the antisymmetry
condition reads
\begin{gather*}
  \ket{\psi(y)} = - E \ket{\psi(-y)},
\intertext{$E$ being the exchange matrix}
  E =
  \begin{pmatrix}
    1 & 0 & 0 & 0 \\
    0 & 0 & 1 & 0 \\
    0 & 1 & 0 & 0 \\
    0 & 0 & 0 & 1 \\
  \end{pmatrix}.
\end{gather*}

\section{Symmetries of the Thirring Quantum Walk}

The Thirring walk manifests some symmetries that allow to simplify the
derivation and the study of the solutions. First of all, as we already
mentioned, one can show that the interaction $V(\chi)$ commutes with the total
number operator. This means that one can study the walk dynamics separately for
each fixed number of particles. We focus here on the two-particle walk $U_2 =
W_2 V_2(\chi)$, where $W_2 = W \otimes W$ and $V_2(\chi) = e^{i\chi\delta_{y,0}
(1-\delta_{a_1,a_2})}$.

Since the interacting walk $U_2$ commutes with the translations in the centre of
mass coordinate $w$, the total momentum is a conserved quantity, so it is
convenient to study the walk parameterized by the total momentum $p$. To this
end we consider the basis $\ket{a_1,a_2}\ket{y}\ket{p}$, so that for fixed
values of $p$ the interacting walk of two particles can be expressed in terms of
a one-dimensional QW $U_2(\chi,p) = W_2(p) V(\chi)$ with a four dimensional
coin:
\begin{gather*}
  W_2(p) = \mu\nu
  \begin{pmatrix}
    \frac{\nu}{\mu} e^{i2p} &
    -i e^{ip} T_y &
    -i e^{ip} \ct T_y &
    -\frac{\mu}{\nu} \\
    -i e^{ip} T_y &
    \frac{\nu}{\mu} T_y^2 &
    -\frac{\mu}{\nu} &
    -i e^{-ip} T_y \\
    -i e^{ip} \ct T_y &
    -\frac{\mu}{\nu} &
    \frac{\nu}{\mu} {\ct T_y}^2 &
    -i e^{-ip} \ct T_y \\
    -\frac{\mu}{\nu} &
    -i e^{-ip} T_y &
    -i e^{-ip} \ct T_y &
    \frac{\nu}{\mu} e^{-i2p}
  \end{pmatrix},
\end{gather*}

Although the range of the variable $p$ is the interval $(-\pi,\pi]$, it is
possible to show that one can restrict the study of the walk to the interval
$[0,\pi/2]$. On the one hand, the two-particle walk transforms unitarily under a
parity transformation in the momentum space. Starting from the single particle
walk, $W(p)$ transforms under a parity transformation as
\begin{gather*}
  W(p) = \sigma_x W(-p) \sigma_x, \qquad p \in (-\pi,\pi],
\end{gather*}
so that for the two-particle walk we have the relation
\begin{equation*}
  W_2(-p,y) = \sigma_x \otimes \sigma_x \, E W_2(p,y) E \,
    \sigma_x \otimes \sigma_x.
\end{equation*}
On the other hand, a translation of $\pi$ of the total momentum $p$ entails that
\begin{gather}
  W_2(p+\pi,y) = \sigma_z \otimes \sigma_z \, W_2(p,y) \,
    \sigma_z \otimes \sigma_z,
\end{gather}
while the interaction term remains unaffected in both cases.

The Thirring walk features also another symmetry that can be exploited to
simplify the derivation of the solutions. It is easy to check that the walk
operator $U_2(p,\chi) = W_2(p) V(\chi)$ commutes with the projector defined by
\begin{equation} \label{eq:interac-proj}
  P \deq
  \begin{pmatrix}
    P_o & 0 & 0 & 0 \\
    0 & P_e & 0 & 0 \\
    0 & 0 & P_e & 0 \\
    0 & 0 & 0 & P_o \\
  \end{pmatrix},
\end{equation}
where $P_e$ and $P_o$ are the projectors on the even and the odd subspaces,
respectively:
\begin{align*}
  & P_e = \sum_{z\in\Z} \ketbra{2z}{2z}, \qquad
  P_o = \sum_{z\in\Z} \ketbra{2z+1}{2z+1}.
\end{align*}
The projector $P$ induces a splitting of the total Hilbert space $\H$
in two subspaces $P\H$ and $(I-P)\H$, with the interaction term acting non trivially only in the subspace $P\H$. In the complementary subspace $(I-P)\H$ the evolution is free for
Fermionic particles. This means that solutions of the free theory are
also solutions of the interacting one, as opposed to the Bosonic case
for which the interaction is non-trivial also in $(I-P)\H$.

\section{Review of the solutions} \label{sec:review-sol}

We focus in this section on the antisymmetric solutions of the Thirring walk
which actually feel the interaction. From the remarks that we have made in the
previous section, such solutions can only be found in the subspace $P\H$.
Formally, we have to solve the eigenvalue equation $PU_2(\chi,p) \ket{\psi} =
e^{-i\omega} \ket{\psi}$, with $\ket{\psi} \in P\H$. Conveniently, we write a
vector $\ket\psi \in P\H$ in the form
\begin{equation}
  \ket\psi = \sum_{z \in \Z}
  \mvec{
    \psi^1(z) \\
    0 \\
    0 \\
    \psi^4(z)
  } \otimes \ket{2z+1} +
  \sum_{z \in \Z}
  \mvec{
    0 \\
    \psi^2(z) \\
    \psi^3(z) \\
    0
  } \otimes \ket{2z},
\end{equation}
and the antisymmetry condition becomes:
\begin{align*}
  & \psi^{1,4}(-z) = - \psi^{1,4}(z-1), \\
  & \psi^{2}(-z)   = - \psi^{3}(z).
\end{align*}
The restriction of the walk to the subspace $P\H$ entails that the eigenvalue
problem is equivalent to the following system of equations:
\begin{equation} \label{eq:rec-th}
\begin{cases}
  e^{-i\omega} \psi^1(z) =
  \nu^2 e^{i2p} \psi^1(z)
  -i\mu\nu e^{ip} e^{i\chi\delta_{z,0}} \psi^2(z)
  -i\mu\nu e^{ip} e^{i\chi\delta_{z,-1}} \psi^3(z+1)
  -\mu^2 \psi^4(z), \\
  e^{-i\omega} \psi^2(z) =
  -i\mu\nu e^{ip} \psi^1(z-1)
  +\nu^2 e^{i\chi\delta_{z,1}} \psi^2(z-1)
  -\mu^2 e^{i\chi\delta_{z,0}} \psi^3(z)
  -i\mu\nu e^{-ip} \psi^4(z-1), \\
  e^{-i\omega} \psi^3(z) =
  -i\mu\nu e^{ip} \psi^1(z)
  -\mu^2 e^{i\chi\delta_{z,0}} \psi^2(z)
  +\nu^2 e^{i\chi\delta_{z,-1}} \psi^3(z+1)
  -i\mu\nu e^{-ip} \psi^4(z), \\
  e^{-i\omega} \psi^4(z) =
  -\mu^2 \psi^1(z)
  -i\mu\nu e^{-ip} e^{i\chi\delta_{z,0}} \psi^2(z)
  -i\mu\nu e^{-ip} e^{i\chi\delta_{z,-1}} \psi^3(z+1)
  +\nu^2 e^{-i2p} \psi^4(z).
\end{cases}
\end{equation}
The most general solution of \cref{eq:rec-th} for $p \not \in
\{0,\pi/2\}$ has two forms:
\begin{gather}
  U_2(\chi,p)\ket{\psi_{\pm\infty}} =
    e^{\pm i 2p} \ket{\psi_{\pm\infty}}, \qquad
  \psi_{\pm\infty}(z) =
  \begin{cases}
    \mvec{
      \zeta_{\pm\infty}\\
      \eta_{\pm\infty}\\
      -\eta_{\pm\infty}\\
      \zeta'_{\pm\infty}
    } \delta_{z,0}, & z \geq 0, \\
    \text{antisymmetrized,} & z < 0,
  \end{cases} \label{eq:inf-sol} \\
\intertext{and}
\begin{aligned}
  & \psi(z) =
  \begin{cases}
    \displaystyle
    \sum_{s,r=\pm} \int_\sS\!\dd k \,
      g_\omega^{sr}(k) \bw^{sr}_k(z), & z > 0, \\
    \text{antisymmetrized,} & z < 0,
  \end{cases} \qquad\quad
  \bw^{sr}_k(z) \deq \mvec{
    \bv_k^{sr,1} e^{-i(2z+1)k} \\
    \bv_k^{sr,2} e^{-i(2z)k} \\
    \bv_k^{sr,3} e^{-i(2z)k} \\
    \bv_k^{sr,4} e^{-i(2z+1)k} \\
  }, \\
  & \psi(0) = \mvec{
    \sum_{s,r=\pm} \int_\sS\!\dd k \,
      g_\omega^{sr}(k) \bv^{sr,1}_k \\
    \xi \\
    -\xi \\
    \sum_{s,r=\pm} \int_\sS\!\dd k \,
      g_\omega^{sr}(k) \bv^{sr,4}_k
  }, \\
\end{aligned} \label{eq:gen-sol} \\
\intertext{with $k = k_R + ik_I$, $\sS \deq \set{k\in\C | k_R \in (-\pi,\pi]}$,
and $g^{sr}_\omega$ satisfying the condition}
  e^{-i\omega} \neq e^{-i\omega_{sr}(p,k)} \implies
    g^{sr}_\omega(k) = 0. \nonumber
\end{gather}
Solving \cref{eq:rec-th} corresponds now to find the function $g_\omega^{sr}$.
Let us now study the equation
\begin{equation*}
  e^{-i\omega_{sr}(p,k)} = e^{-i\omega}.
\end{equation*}
Since $e^{-i\omega_{sr}(p,k)}$ has to be an eigenvalue of $U_2(\chi,p)$,
$\omega_{sr}(p,k)$ must be real and thus $k \in \Gamma_f$ or $k \in \Gamma_l$
with $l = 0,\pm 1,2$, so we conveniently define the sets:
\begin{align*}
  & \Omega_f^{sr} \deq \Set{ e^{-i\omega_{sr}(p,k)} | k \in \Gamma_{f} }, \qquad
    \Omega^{sr}_{l} \deq \Set{ e^{-i\omega_{sr}(p,k)} | k \in \Gamma_l }, \\
  & \Gamma_f \deq \Set{ k \in \sS | k_R \in (-\pi,\pi] }, \qquad
  \Gamma_{l} \deq \Set{ k \in \sS | k_R = l \frac{\pi}{2} }, \qquad
    l = 0,\pm 1,2.
\end{align*}
It is easy to see that $\Omega_f^{sr} \cap \Omega_l^{sr} = \emptyset$ for all
$s$, $r$ and $l$, and the range of the function $e^{-i\omega_{sr}(p,k)}$ covers
the entire unit circle except for the points $e^{\pm i 2p}$. Therefore, we can
discuss separately the case $e^{-i\omega} \in \Omega_f^{sr}$ and the case
$e^{-i\omega} \in \Omega_l^{sr}$. A solution with $e^{-i\omega} = e^{\pm i 2p}$
actually exists, corresponding to the function of \cref{eq:inf-sol}, and it will
be discussed in \cref{sec:inf-bound}.

Let us start with the case $e^{-i\omega} \in \Omega_f^{sr}$ which will lead to
the characterization of the continuous spectrum of the Thirring walk
$U_2(\chi,p)$ and of the scattering solutions.

\begin{figure}[t]
  \centering
  \includegraphics[width=0.5\textwidth]{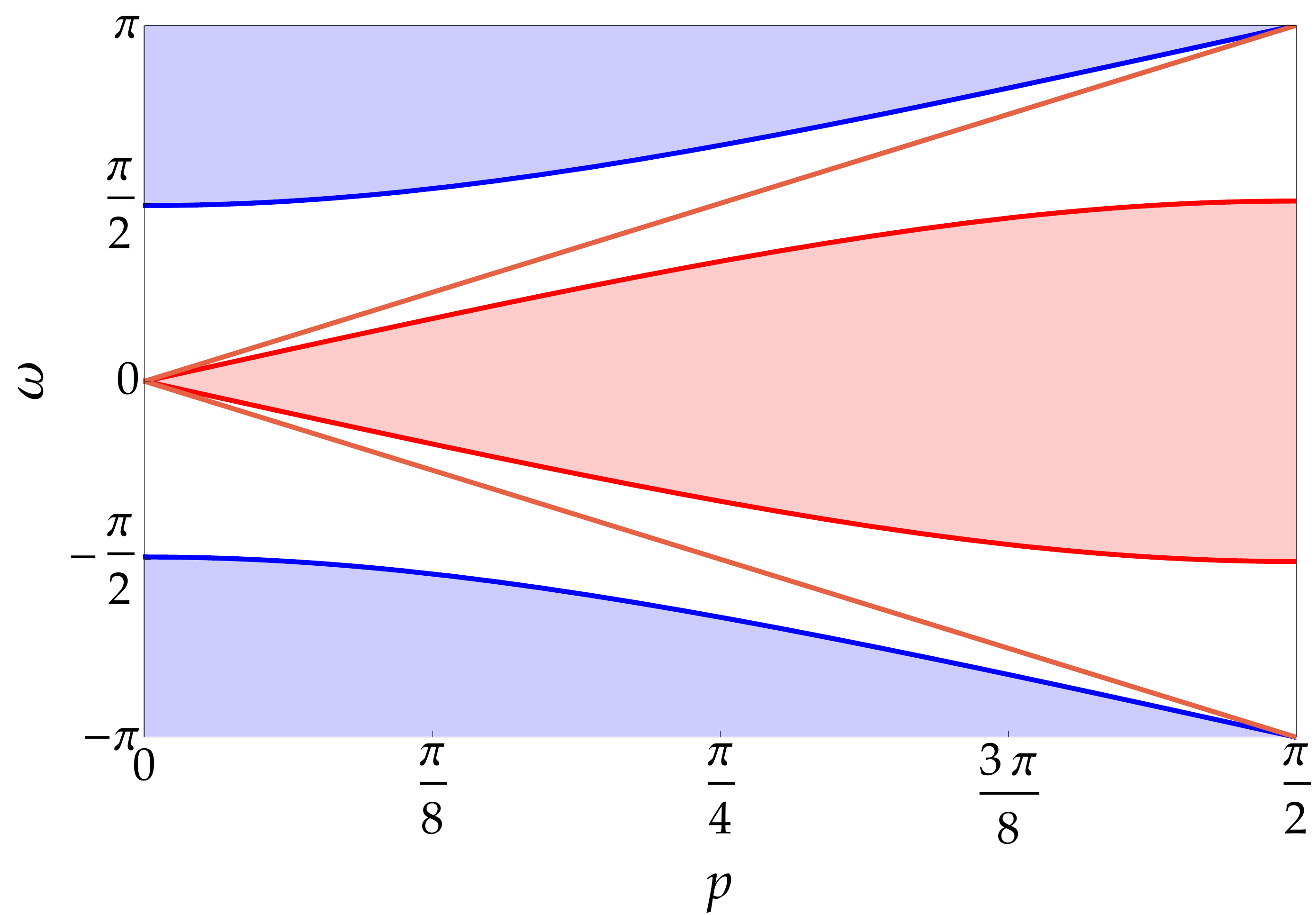}
  \caption{Continuous spectrum of the two-particle walk as a function of the
  total momentum $p \in [0,\pi/2]$ with mass parameter $m=0.7$. The continuous
  spectrum is the same as in the free case. The solid blue curves are described
  by the functions $\omega = \pm 2\omega(p)$, and the red ones by $\omega =
  \pm(\pi-2\Arccos(n\sin p))$. As one can notice, the light-red lines $\omega = \pm
  2p$ lie entirely in the gaps between the solid curves, highlighting the fact
  that $e^{\pm i2p}$ is not in the range of $e^{-i\omega_{sr}(p,k)}$ for $p \neq
  0, \, \pi/2$ (see text).}
  \label{fig:bande-free}
\end{figure}

\subsection{Scattering solutions}

\begin{figure}[t]
  \centering
  \includegraphics[width=0.6\textwidth]{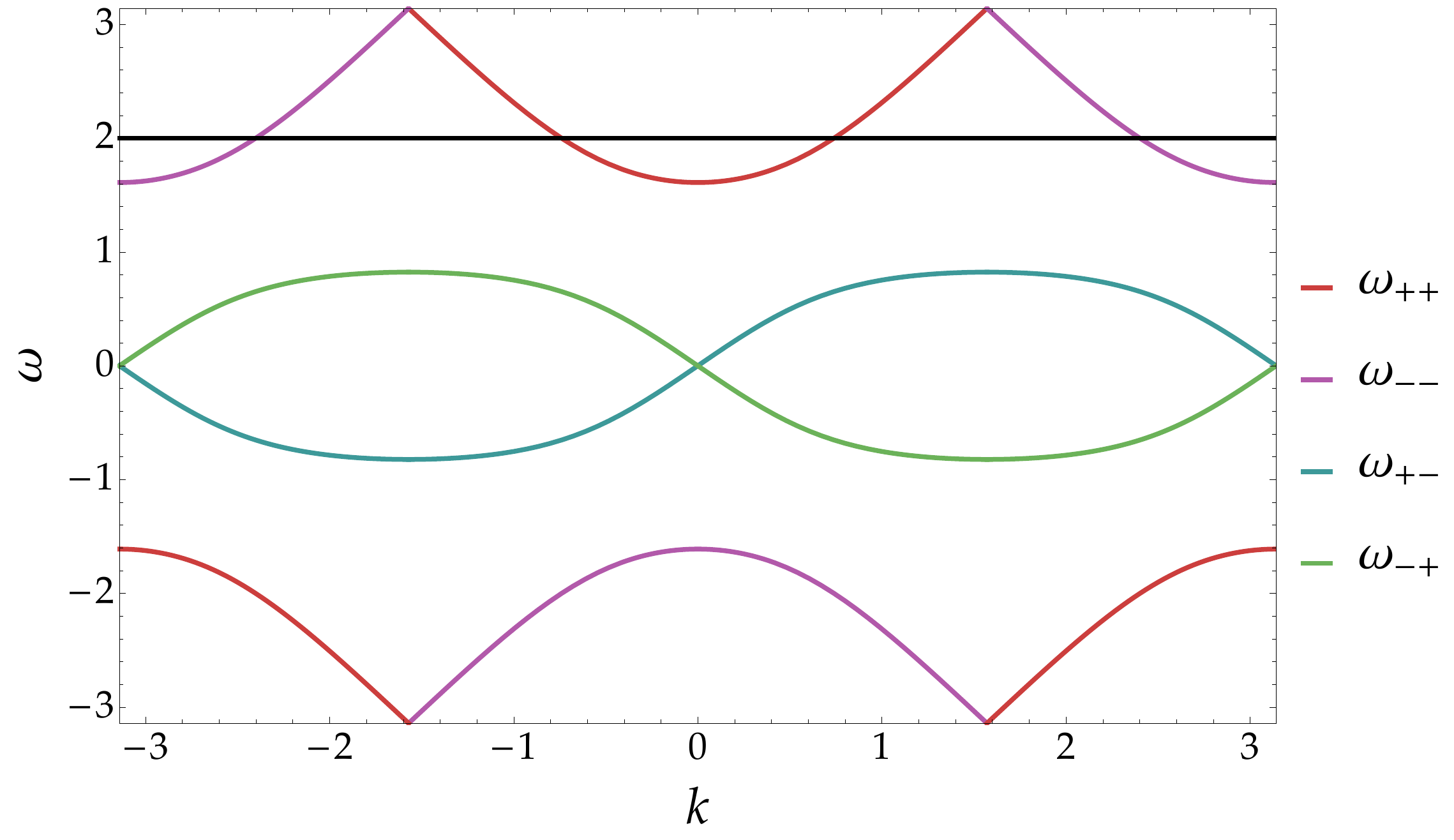}
  \caption{Spectrum of the walk for $m=0.6$ and $p=\pi/6$ as a function of $k$.
  The colours highlight the different ranges of eigenvalues corresponding to the
  dispersion relation $\omega_{sr}(p,k)$. The range of $\omega_{sr}(p,k)$ is
  understood to be computed $\bmod(2\pi)$. One can notice that there are four
  values of the relative momentum $k$ having the same value of the dispersion
  relation ($\omega = 2$ in the figure). This is in contrast to the Hamiltonian
  model for which there are only two solutions.}
  \label{fig:spectrum-sol}
\end{figure}

In this section we assume $p \not \in \{0,\, \pi/2\}$ with $e^{-i\omega} \in
\Omega_f^{sr}$. This implies that $e^{-i\omega} \neq e^{\pm i 2p}$: indeed, as
one can notice from \cref{fig:bande-free}, the lines $\omega = \pm 2p$ lie
entirely in the gaps between the curves $\omega = \pm 2\omega(p)$ and $\omega =
\pm(\pi-2\Arccos(n\sin p))$. The solution is thus the one given in
\cref{eq:gen-sol}. One can  prove that $\Omega_f^{++} = \Omega_f^{--}$ and
$\Omega_f^{+-} = \Omega_f^{-+}$. Furthermore, as one can notice from
\cref{fig:spectrum-sol}, there are four values of the triple $(s,r,k)$ such that
$e^{-i\omega_{sr}(p,k)} = e^{-i\omega}$ for a given value of $e^{-i\omega}$: if
the triple $(+,+,k)$ is a solution, so are $(+,+,\pi-k)$, $(-,-,-k)$ and
$(-,-,k-\pi)$; and if $(+,-,k)$ is a solution, then also $(+,-,\pi-k)$,
$(-,+,-k)$ and $(-,+,k-\pi)$ are solutions. This result greatly simplifies
\cref{eq:gen-sol}. Indeed the sum over $s,r$ and the integral over $k$ reduces
to the sum of four terms:
\begin{equation} \label{eq:sc-ansatz}
\begin{aligned}
  & \psi^{\pm,1}_k(z) \deq
    (\alpha_k^\pm v^{+\pm,1}_k +
    \delta_k^\pm v^{-\mp,1}_{k-\pi} ) e^{-i(2z+1)k} -
    (\beta_k^\pm v^{\pm+,1}_{-k} +
    \gamma_k^\pm v^{\mp-,1}_{\pi-k} ) e^{i(2z+1)k}, \qquad
    & & z \geq 0, \\
  & \psi^{\pm,2}_k(z) \deq
    (\alpha_k^\pm v^{+\pm,2}_k -
    \delta_k^\pm v^{-\mp,2}_{k-\pi} ) e^{-i2zk} -
    (\beta_k^\pm v^{\pm+,2}_{-k} -
    \gamma_k^\pm v^{\mp-,2}_{\pi-k} ) e^{i2zk}, \qquad
    & & z > 0, \\
  & \psi^{\pm,3}_k(z) \deq
    (\alpha_k^\pm v^{+\pm,3}_k -
    \delta_k^\pm v^{-\mp,3}_{k-\pi} ) e^{-i2zk} -
    (\beta_k^\pm v^{\pm+,3}_{-k} -
    \gamma_k^\pm v^{\mp-,3}_{\pi-k} ) e^{i2zk}, \qquad
    & & z > 0, \\
  & \psi^{\pm,4}_k(z) \deq
    (\alpha_k^\pm v^{+\pm,4}_k +
    \delta_k^\pm v^{-\mp,4}_{k-\pi} ) e^{-i(2z+1)k} -
    (\beta_k^\pm v^{\pm+,4}_{-k} +
    \gamma_k^\pm v^{\mp-,4}_{\pi-k} ) e^{i(2z+1)k},
    & & z \geq 0, \\
  & \psi^{\pm,2}_k(0) = -\psi^{\pm,3}_k(0) \deq \xi.
\end{aligned}
\end{equation}

%
As we will see, the original problem can be simplified in this way to an
algebraic problem with a finite set of equations. We remark that the fact that
the equation $e^{-i\omega_{sr}(p,k)} = e^{-i\omega}$ has a finite number of
solutions is a consequence of the fact that we are considering a model in one
spatial dimension. However, in analogous one-dimensional Hamiltonian models (\eg
the Hubbard model) the degeneracy of the eigenvalues is two.

Let us consider for the sake of simplicity the solution of the kind
$\psi_k^{+,j}(z)$, since the other one can be analysed in a similar way. Using
the notation of \cref{app:eigen-free}, \cref{eq:sc-ansatz} reduces to
the expressions (dropping the $+$ superscript)
\begin{equation} \label{eq:ansatz}
\begin{aligned}
  & \psi^{1}_k(z) =
    a [ \lambda e^{-i(2z+1)k} -
    \rho e^{i(2z+1)k} ], \\
  & \psi^{2}_k(z) =
    \lambda b e^{-i2zk} -
    \rho c e^{i2zk}, \\
  & \psi^{3}_k(z) =
    \lambda c e^{-i2zk} -
    \rho b e^{i2zk}, \\
  & \psi^{4}_k(z) =
    d [ \lambda e^{-i(2z+1)k} -
    \rho e^{i(2z+1)k} ], \\
  & \lambda \deq \alpha_k + \delta_k, \qquad
  \rho \deq \beta_k + \gamma_k, \\
  & \psi^2_k(0) = \xi.
\end{aligned}
\end{equation}
We notice that now the number of unknown parameters is further reduced to three,
namely $\lambda$, $\rho$, and $\xi$. Clearly, one of the parameters can be fixed
by choosing arbitrarily the normalization. From now on we fix $\lambda = 1$ and
define $T_+ \deq \rho$. \cref{eq:ansatz} has to satisfy the recurrence relations
of \cref{eq:rec-th} for $z=0$ and $z=1$, while for $z>1$ it is automatically
satisfied. For $z=0$, \cref{eq:rec-th} becomes
\begin{numcases}{} \label{eq:rec-th-0}
  e^{-i\omega} \psi^1_k(0) =
  \nu^2 e^{i2p} \psi^1_k(0)
  -i\mu\nu e^{ip} e^{i\chi} \xi
  -i\mu\nu e^{ip} \psi^3_k(1)
  -\mu^2 \psi^4_k(0),
  \label{eq:rec-1} \\
  e^{-i\omega} \xi =
  i\mu\nu e^{ip} \psi^1_k(0)
  -\nu^2 \psi^3_k(1)
  -\mu^2 e^{i\chi} \xi
  +i\mu\nu e^{-ip} \psi^4_k(0),
  \label{eq:rec-2} \\
  -e^{-i\omega} \xi =
  -i\mu\nu e^{ip} \psi^1_k(0)
  -\mu^2 e^{i\chi} \xi
  +\nu^2 \psi^3_k(1)
  -i\mu\nu e^{-ip} \psi^4_k(0),
  \label{eq:rec-3} \\
  e^{-i\omega} \psi^4_k(0) =
  -\mu^2 \psi^1_k(0)
  -i\mu\nu e^{-ip} e^{i\chi} \xi
  -i\mu\nu e^{-ip} \psi^3_k(1)
  +\nu^2 e^{-i2p} \psi^4_k(0).
  \label{eq:rec-4}
\end{numcases}

\noindent Starting from Eq.~\eqref{eq:rec-1}, we can notice that
$\nu^2 e^{i2p} a
-i\mu\nu e^{ip} e^{ik} b
-i\mu\nu e^{ip} e^{-ik} c
-\mu^2 d = e^{-i\omega} a$,
where we employed the notation of \cref{app:eigen-free}, so that we obtain $\xi
= e^{-i\chi} (b - T_+ c)$. We can then substitute this expression in
Eq.~\eqref{eq:rec-3} and use the relations
\begin{gather*}
  -i\mu\nu e^{ip} e^{ik} a + \nu^2 e^{i2k} b
  -\mu^2 c - i\mu\nu e^{-ip} e^{ik} d = e^{-i\omega} b, \\
  -i\mu\nu e^{ip} e^{-ik} a - \mu^2 b
  +\nu^2 e^{-i2k} c - i\mu\nu e^{-ip} e^{-ik} d = e^{-i\omega} c, \\
\intertext{to obtain the expression}
  e^{-i\chi} (b - T_+ c) = T_+ b - c,
\end{gather*}
and thus
\begin{equation} \label{eq:tr-cof-plus}
  T_+ = \frac{c + e^{-i\chi}b}{b + e^{-i\chi}c}
  = \frac{g_+(p+k) + e^{-i\chi} g_+(p-k)}{g_+(p-k) + e^{-i\chi} g_+(p+k)}.
\end{equation}
For these values of $\xi$ and $T_+$ one can verify that \cref{eq:rec-th} is
satisfied also for $z=1$, thus concluding the derivation. For the solution of
the kind $\psi_k^{-,j}(z)$ we can follow a similar reasoning, obtaining the
analogous quantity $T_-$:
\begin{equation} \label{eq:tr-cof-minus}
  T_- \deq \frac{g_+(p+k) + e^{-i\chi} g_-(p-k)}{g_-(p-k) +
    e^{-i\chi} g_+(p+k)}.
\end{equation}
It is worth noticing that $T_\pm$ is of unit modulus for $k \in (-\pi,\pi]$.

\begin{figure}[t]
  \centering
  \includegraphics[width=0.48\textwidth]{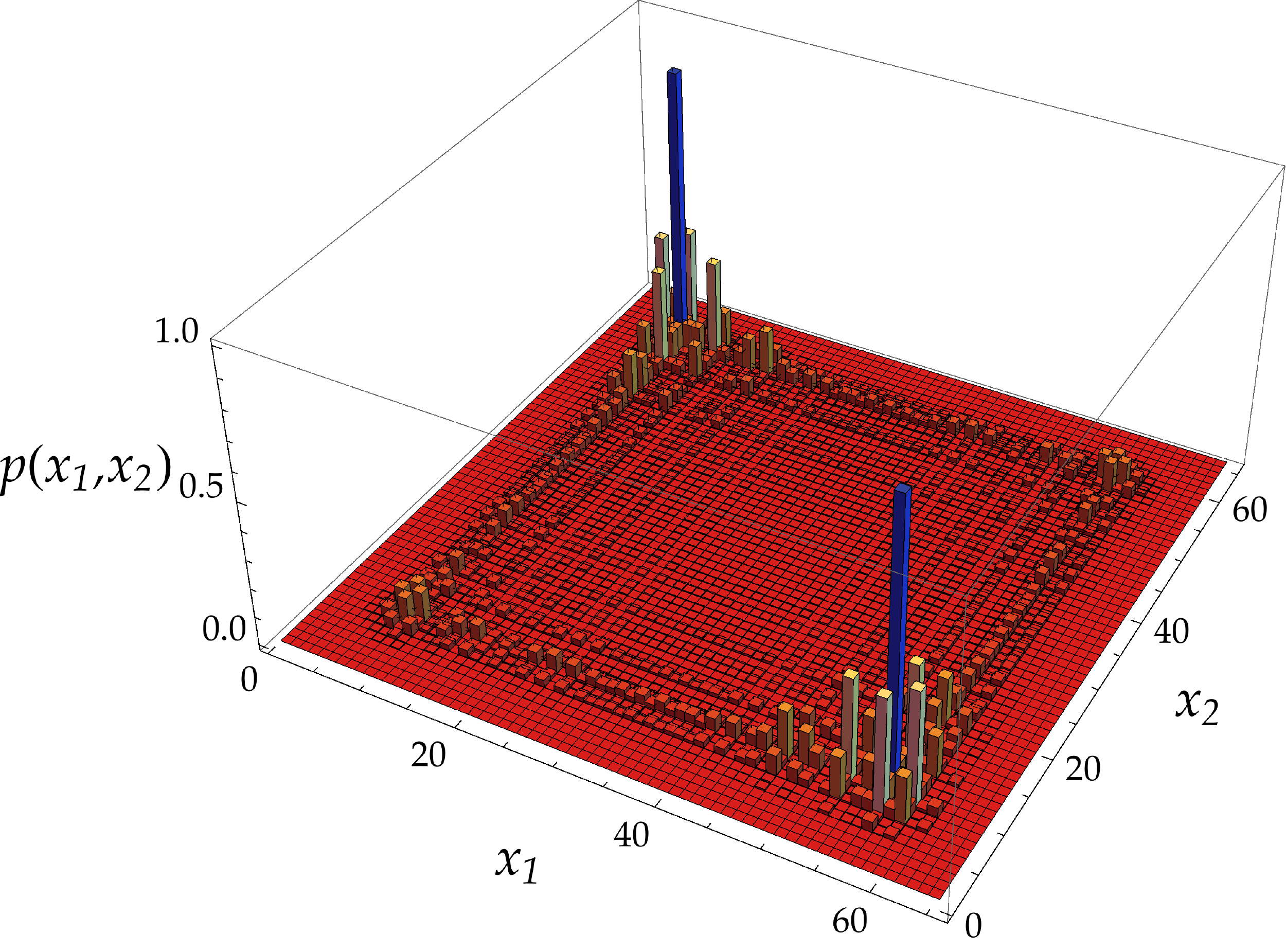}
  \quad
  \includegraphics[width=0.48\textwidth]{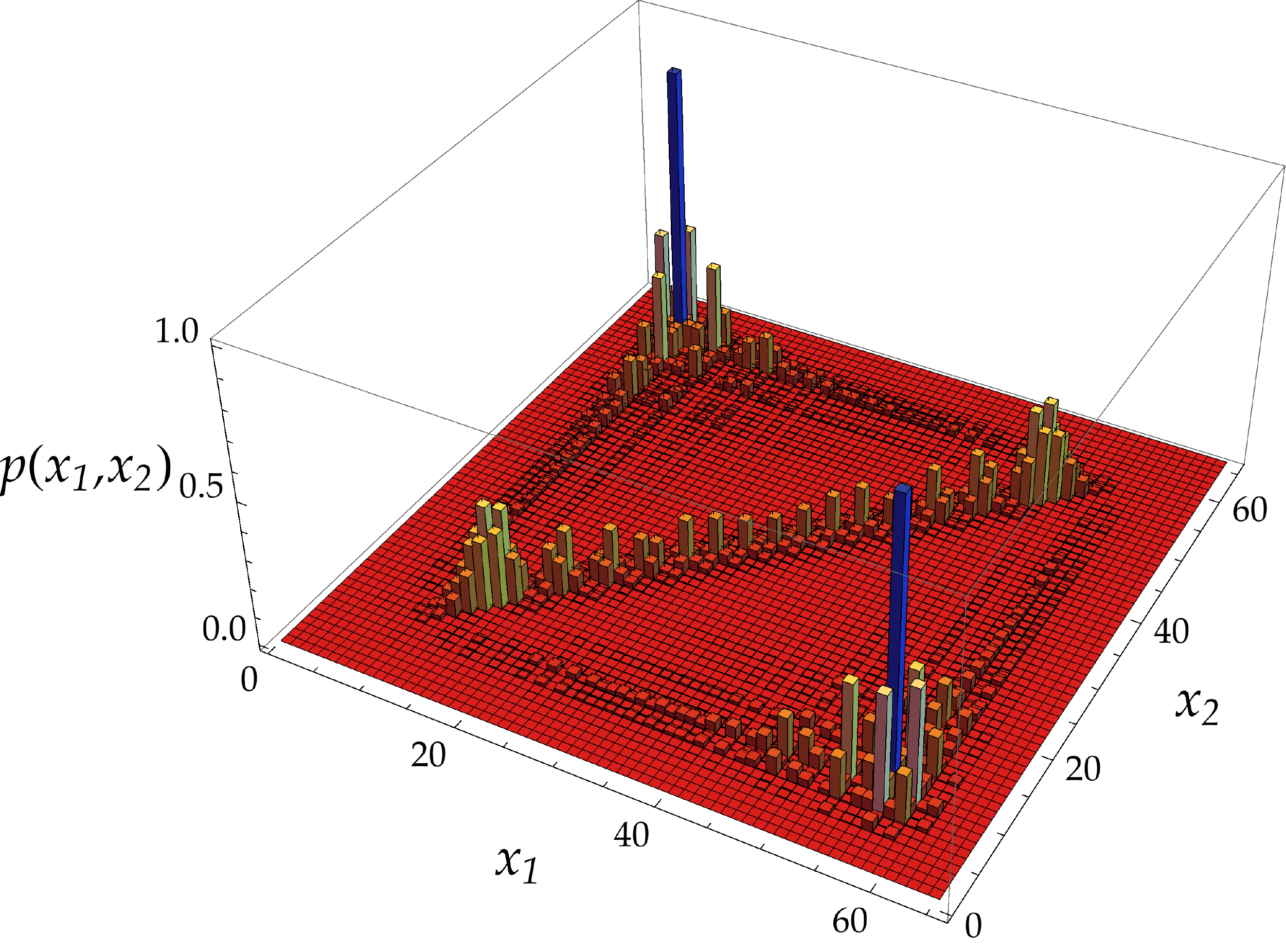}
  \caption{We show for comparison the free evolution (left panel) and the
  interacting one (right panel) highlighting the appearance of bound states
  components along the diagonal, namely when the two particles are at the same
  site (\ie $x_1 = x_2$), where $x_1$ and $x_2$ denote the positions of the two
  particles. The plots show the probability distribution $p(x_1,x_2)$ in
  position space after $t = 32$ time-steps.  The chosen value of the mass
  parameter is $m=0.6$ and the coupling constant is $\chi=\pi/2$. The two
  particles are initially prepared in a singlet state located at the origin.}
  \label{fig:loc-bound}
\end{figure}

The final form of the solution results to be
\begin{equation} \label{eq:sol}
\begin{aligned}
  & \psi_k^{\pm,1}(z) = (v^{+\pm,1}_k + v^{-\mp,1}_{k-\pi}) e^{-i(2z+1)k}
    - T_\pm (v^{\pm+,1}_{-k} + v^{\mp-,1}_{\pi-k}) e^{i(2z+1)k}, \\
  & \psi_k^{\pm,2}(z) = e^{-i\chi\delta_{z,0}} \left[
    (v^{+\pm,2}_k - v^{-\mp,2}_{k-\pi}) e^{-i2zk}
    - T_\pm (v^{\pm+,2}_{-k} - v^{\mp-,2}_{\pi-k}) e^{i2zk} \right], \\
  & \psi_k^{\pm,3}(z) = (v^{+\pm,3}_k - v^{-\mp,3}_{k-\pi}) e^{-i2zk}
    - T_\pm (v^{\pm+,3}_{-k} - v^{\mp-,3}_{\pi-k}) e^{i2zk}, \\
  & \psi_k^{\pm,4}(z) = (v^{+\pm,4}_k + v^{-\mp,4}_{k-\pi}) e^{-i(2z+1)k}
    - T_\pm (v^{\pm+,4}_{-k} + v^{\mp-,4}_{\pi-k}) e^{i(2z+1)k},
\end{aligned}
\end{equation}
which in terms of the relative coordinate $y$ can be written as
\begin{equation*}
  \psi_k^{\pm}(y) =
  \begin{cases}
    e^{-i\chi\delta_{z,0}\delta_{j,2}} \left[
      (v^{+\pm}_k + v^{-\mp}_{k-\pi}) e^{-iky}
        - T_\pm (v^{\pm+}_{-k} + v^{\mp-}_{\pi-k}) e^{iky}
    \right], & y \geq 0, \\
    \text{antisymmetrized,} & y < 0.
  \end{cases}
\end{equation*}
We can interpret such a solution as a scattering of plane waves for which the
coefficient $T_\pm$ plays the role of the transmission coefficient. Being the
total momentum a conserved quantity, the two particles can only exchange their
momenta, as expected from a theory in one-dimension. Furthermore, for each value
$k$ of the relative momentum, the two particles can also acquire an additional
phase of $\pi$. As the interaction is a compact perturbation of the free
evolution, the continuous spectrum is the same as that of the free walk.
\cref{eq:sol} provides the generalized eigenvector if $U_2(\chi,p)$
corresponding to the continuous spectrum $\sigma_c = \Omega_f^{++} \cup
\Omega_f^{+-}$.

\subsection{Bound states} \label{sec:bound}

\begin{figure}[t]
  \centering
  \includegraphics[width=0.6\textwidth]{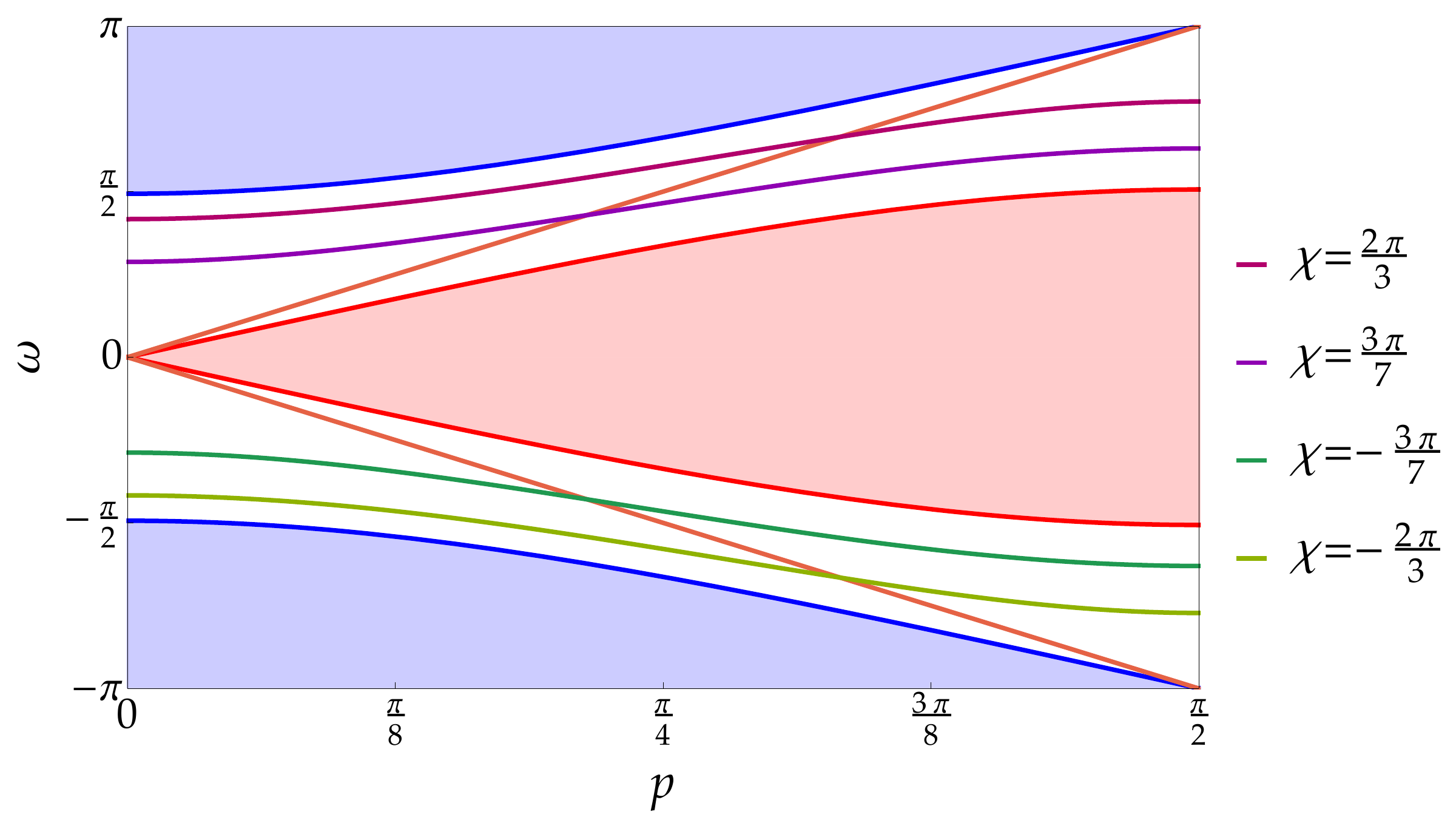}
  \caption{Complete spectrum of the two-particle Thirring walk as a function of
  the total momentum $p$ with mass parameter $m=0.7$. The continuous spectrum is
  as in \cref{fig:bande-free}. The solid lines in the gaps show the point
  spectrum for different values of the coupling constant: from top to bottom,
  $\chi = 2\pi/3,\, 3\pi/7,\, -3\pi/7,\, -2\pi/3$. It is worth noticing that for
  each pair $(\chi,p)$ there is only one value in the discrete spectrum. The
  light-red lines $\omega = \pm 2p$ intersect the curves of the discrete
  spectrum for $p=\chi/2$.}
  \label{fig:bande}
\end{figure}

In the previous section, we derived the solutions in the continuous spectrum,
which can be interpreted as scattering plane waves in one spatial dimension. We
seek now the solutions corresponding to the discrete spectrum, namely solutions
with eigenvalue in any one of the sets $\Omega_l^{sr}$. The derivation of the
solution follows similar steps as for the scattering solutions. In particular,
the degeneracy in $k$ is the same: there are four solutions to the equation
$e^{-i\omega_{sr}(p,k)} = e^{-i\omega}$ even in this case, as proved in
Ref.~\cite{PhysRevA.97.032132}. Therefore the general form of the solution in this
case can be written again as in \cref{eq:sc-ansatz} and, following the same
reasoning, one obtains the same set of solutions as in \cref{eq:sol}. At this
stage we did not imposed that the solution is a proper eigenvector in the
Hilbert space $\H$. To this end, we have to set $T_\pm = 0$ to eliminate the
exponentially-divergent terms in \cref{eq:sol}. As one can prove, the equation
$T_\pm = 0$ has only one solution for fixed values of $\chi$ and $p$. More
precisely, there is a unique $k \in \Gamma_0 \cup \Gamma_{-1} \cup \Gamma_1 \cup
\Gamma_2$, with $k_I < 0$ and $e^{i\chi} \not \in \{1,-1\}$, such that either
$T_+ = 0$ or $T_- = 0$.


In other words, for each pair of values $(\chi,p)$ the walk $U_2(p)$ has one and
only eigenvector corresponding to an eigenvalue in the point spectrum. Such
eigenvector can be written as
\begin{equation} \label{eq:sol-bound}
\begin{aligned}
  & \psi_{\tilde k}^{1}(z) = (v^{+\pm,1}_{\tilde k} + v^{-\mp,1}_{\tilde k-\pi}) e^{-i(2z+1)\tilde k}, \\
  & \psi_{\tilde k}^{2}(z) = e^{-i\chi\delta_{z,0}} \left[
    (v^{+\pm,2}_{\tilde k} - v^{-\mp,2}_{\tilde k-\pi}) e^{-i2z\tilde k} \right], \\
  & \psi_{\tilde k}^{3}(z) = (v^{+\pm,3}_{\tilde k} - v^{-\mp,3}_{\tilde k-\pi}) e^{-i2z\tilde k}, \\
  & \psi_{\tilde k}^{4}(z) = (v^{+\pm,4}_{\tilde k} + v^{-\mp,4}_{\tilde k-\pi}) e^{-i(2z+1)\tilde k},
\end{aligned}
\end{equation}
where $\tilde k$ is the solution of $T_+ = 0$ or $T_- = 0$ and $\pm$ chosen
accordingly. More compactly, in the $y$ coordinate, the solution can be written
as
\begin{equation*}
  \psi_{\tilde k}(y) =
  \begin{cases}
    e^{-i\chi\delta_{z,0}\delta_{j,2}} \left[
    (v^{+\pm}_{\tilde k} + v^{-\mp}_{\tilde k-\pi})
      e^{-i\tilde k y} \right], & y \geq 0, \\
    \text{antisymmetrized,} & y < 0.
  \end{cases}
\end{equation*}

Referring to \cref{fig:loc-bound}, we show the evolution of two particles
initially prepared in a singlet state localized at the origin. From the figure
one can appreciate the appearance of the bound state component which has
non-vanishing overlapping with the initial state. The bound state, being
exponentially decaying in the relative coordinate $y$, is localized on the
diagonal of the plot, that is when the two particles lie at the same point.

\begin{figure}[t]
  \centering
  \includegraphics[width=0.48\textwidth]{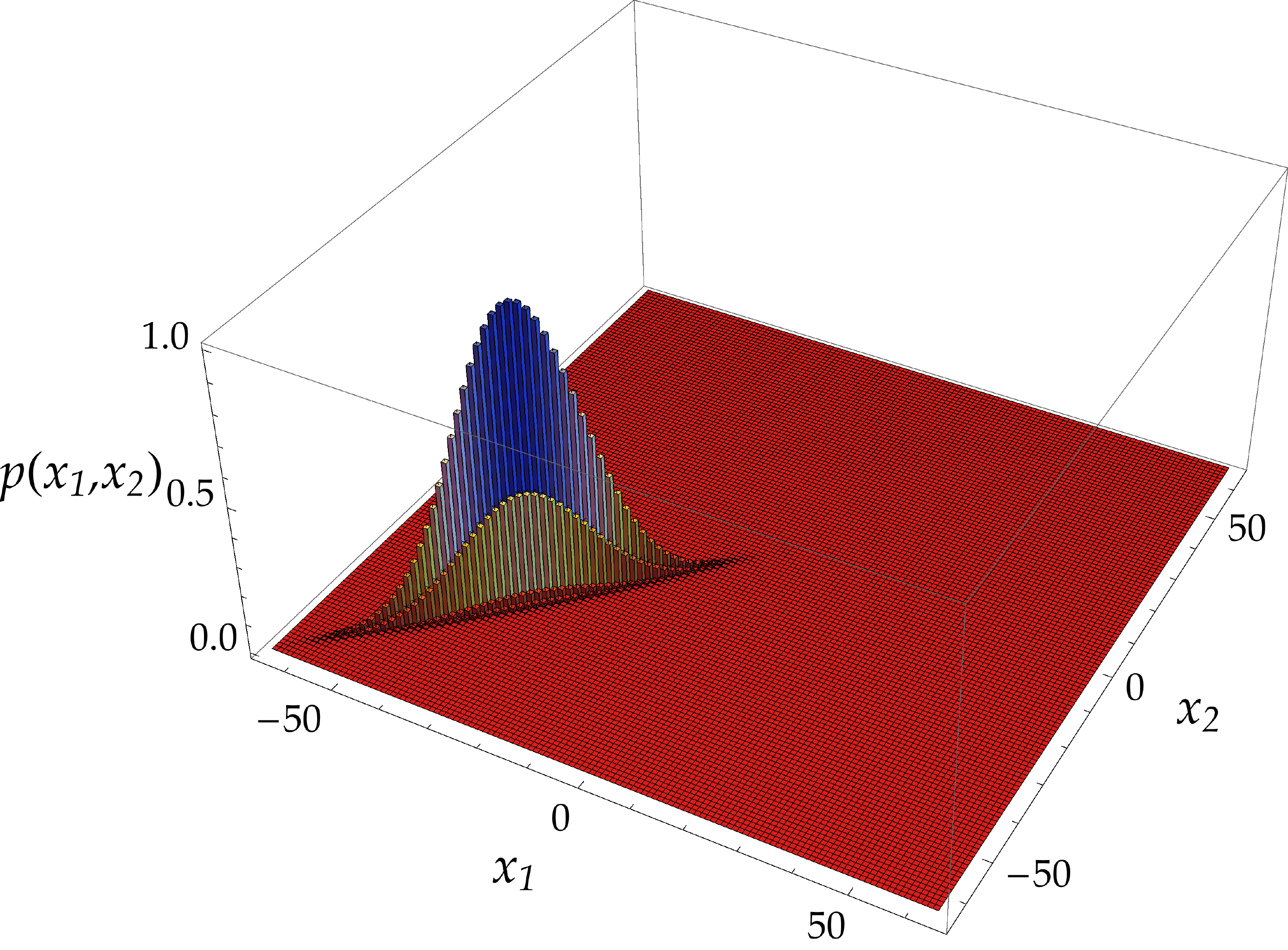}
  \quad
  \includegraphics[width=0.48\textwidth]{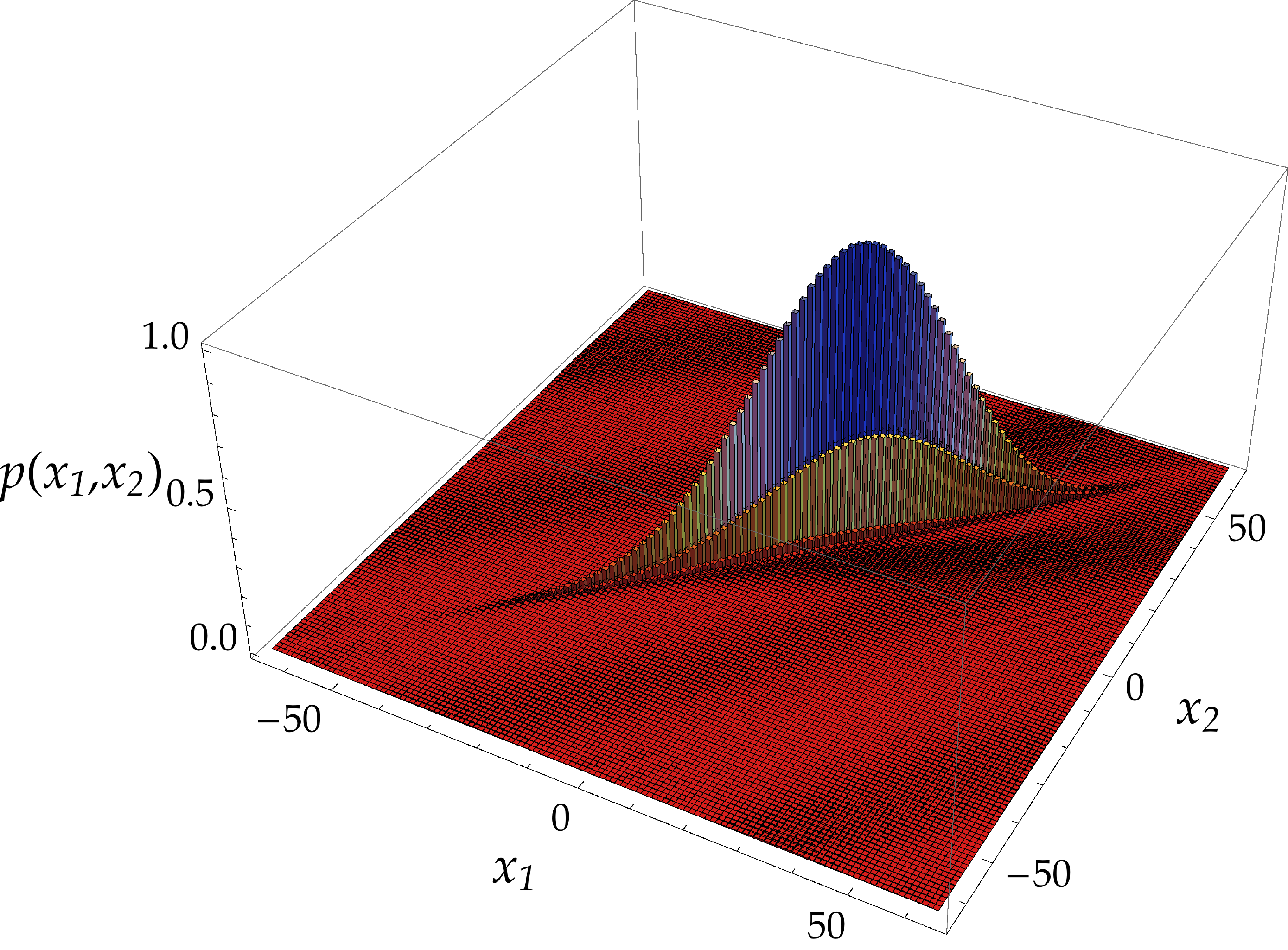}
  \caption{We show the evolution of a bound state of the two particles peaked
  around the value of the total momentum $p=0.035\pi$. The mass paramater is
  $m=0.6$ and the coupling constant $\chi=0.2\pi$. On the left is depicted the
  probability distribution of the initial state and on the right the that of the
  evolved state after $t=128$ time-steps. One can notice that in the relative
  coordinate $x_1-x_2$ the probability distribution remains concentrated on the
  diagonal, highlighting that the two particles are in a bound state. The
  diffusion of the state happens only in the centre of mass coordinate.}
  \label{fig:bound-state}
\end{figure}

In \cref{fig:bound-state} is depicted the probability distribution of the bound state corresponding
to choice of the parameters $\chi=0.2\pi$ and $p=0.035\pi$. The plot highlights the exponential decay of the tails, which is the characterizing feature of the bound state.

\subsection{Solution for \protect{$e^{-i\omega} = e^{\pm i2p}$}}
\label{sec:inf-bound}

So far we have studied proper eigenvectors which decay exponentially as the two
particles are further apart. However, the previous analysis failed to cover the
particular case when $e^{-i\omega} = e^{\pm i2p}$, since the range of
$e^{-i\omega_{sr}(p,k)}$ does not include the two points of the unit circle
$e^{\pm i2p}$.

We study now the solutions with $e^{-i\omega} = e^{\pm i2p}$ having the form
given in \cref{eq:inf-sol}. One can prove that such solutions are non-vanishing
only for $z=0$ on $P\H$, namely we look for a solution of the form
\begin{gather} \label{eq:bound-loc}
  \ket{\psi} =
    \mvec{
      -\zeta\\
      0\\
      0\\
      -\zeta'
    } \otimes \ket{-1} +
    \mvec{
      0\\
      \eta\\
      -\eta\\
      0
    } \otimes \ket{0} +
    \mvec{
      \zeta\\
      0\\
      0\\
      \zeta'
    } \otimes \ket{1}.
\end{gather}
Subtracting the first and the last equations of \eqref{eq:rec-th} using
\eqref{eq:bound-loc}, we obtain the following equation:
\begin{equation} \label{eq:bound-loc-eigenvalues}
  (e^{-i\omega} - e^{i2p}) \zeta = e^{i2p} (e^{-i\omega} - e^{-i2p}) \zeta'.
\end{equation}
If both $\zeta$ and $\zeta'$ are non-zero, one can prove that a solution does
not exist and thus we have to consider the two cases $\zeta = 0$ and $\zeta' =
0$ separately. Starting from $\zeta' = 0$, \cref{eq:bound-loc-eigenvalues}
imposes that $e^{-i\omega} = e^{i2p}$, meaning that if a solution exists in this
case, it is an eigenvector corresponding to the eigenvalue $e^{i2p}$. From the
second equation of \eqref{eq:rec-th} we obtain the relation
\begin{equation*}
  (1 - \mu^2 e^{i(\chi-2p)}) \eta = i\mu\nu e^{-ip} \zeta
\end{equation*}
and, using the first equation of \eqref{eq:rec-th}, it turns out that a solution
exists only if $e^{i\chi} = e^{i2p}$, as expected since otherwise we would have
been in the case of \cref{sec:bound} would hold. The other case, namely
$e^{-i\omega} = e^{-i2p}$, can be studied analogously. Let us, then, denote as
$\ket{\psi_{\pm\infty}}$ such proper eigenvectors with eigenvalue $e^{\pm i2p}$
for $\chi = e^{\pm i2p}$ and, choosing $\eta = \frac{\mu}{\nu}$ as the value for
the free parameter $\eta$, we obtain the following expression for
$\ket{\psi_{\pm\infty}}$:
\begin{gather*}
  \ket{\psi_{\pm\infty}} =
    ie^{\pm ip}
    \mvec{\frac{1\pm 1}{2}\\0\\0\\-\frac{-1\pm 1}{2}} \otimes \ket{-1} +
    \mvec{0\\\frac{\mu}{\nu}\\-\frac{\mu}{\nu}\\0} \otimes \ket{0} +
    ie^{\pm ip}
    \mvec{-\frac{1\pm 1}{2}\\0\\0\\\frac{-1\pm 1}{2}} \otimes \ket{1}.
\end{gather*}
Such solutions provide a special case of molecule states (namely, proper
eigenvectors of $U_2(\chi,p)$), being localized on few sites, and differ from
the previous solutions showing an exponential decay in the relative coordinate.

\subsection{Solutions for \protect{$p \in \{0,\, \pi/2\}$}}
\label{sec:quit-bound}

\begin{figure}[t]
  \centering
  \includegraphics[width=0.48\textwidth]{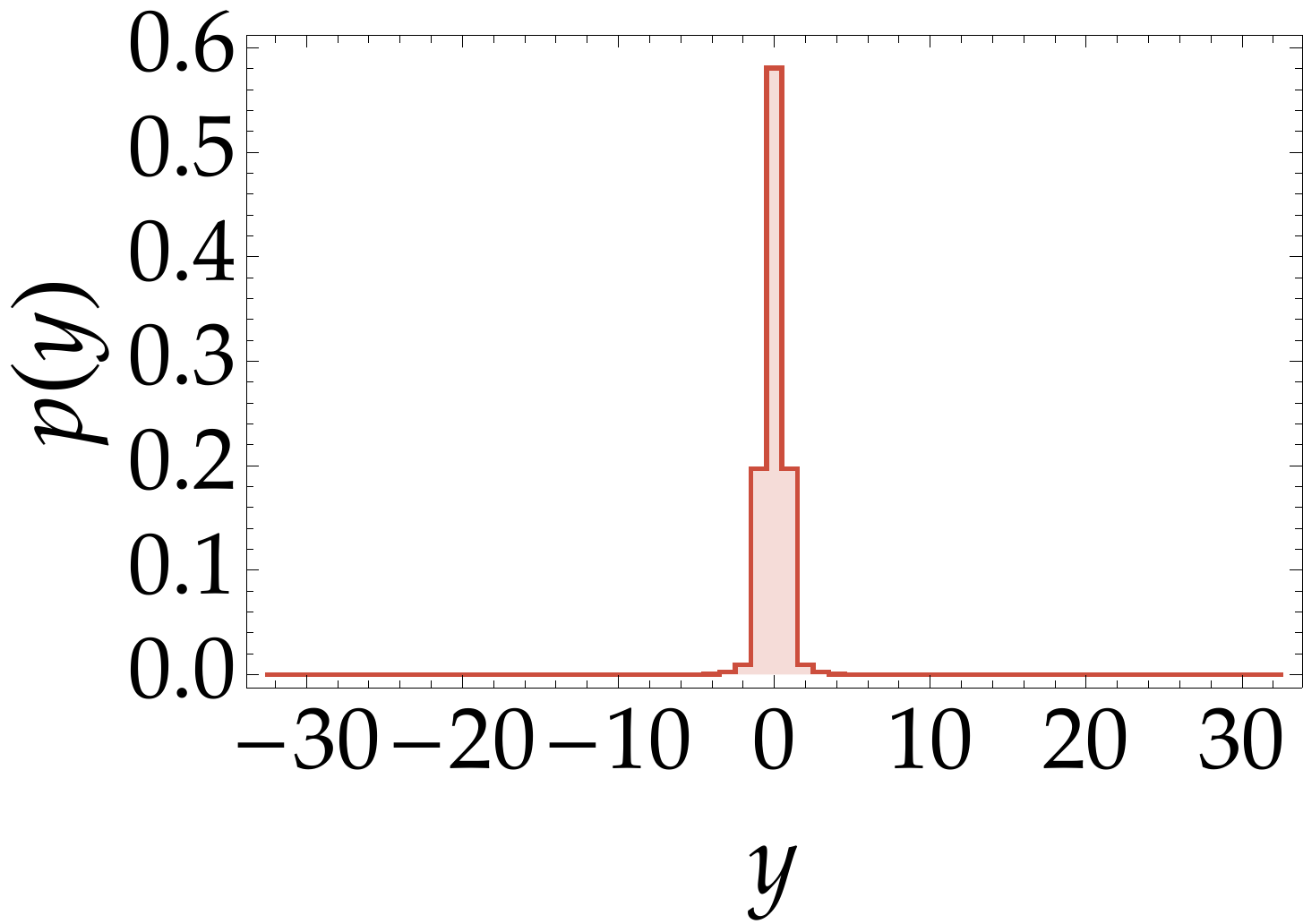}
  \quad
  \includegraphics[width=0.48\textwidth]{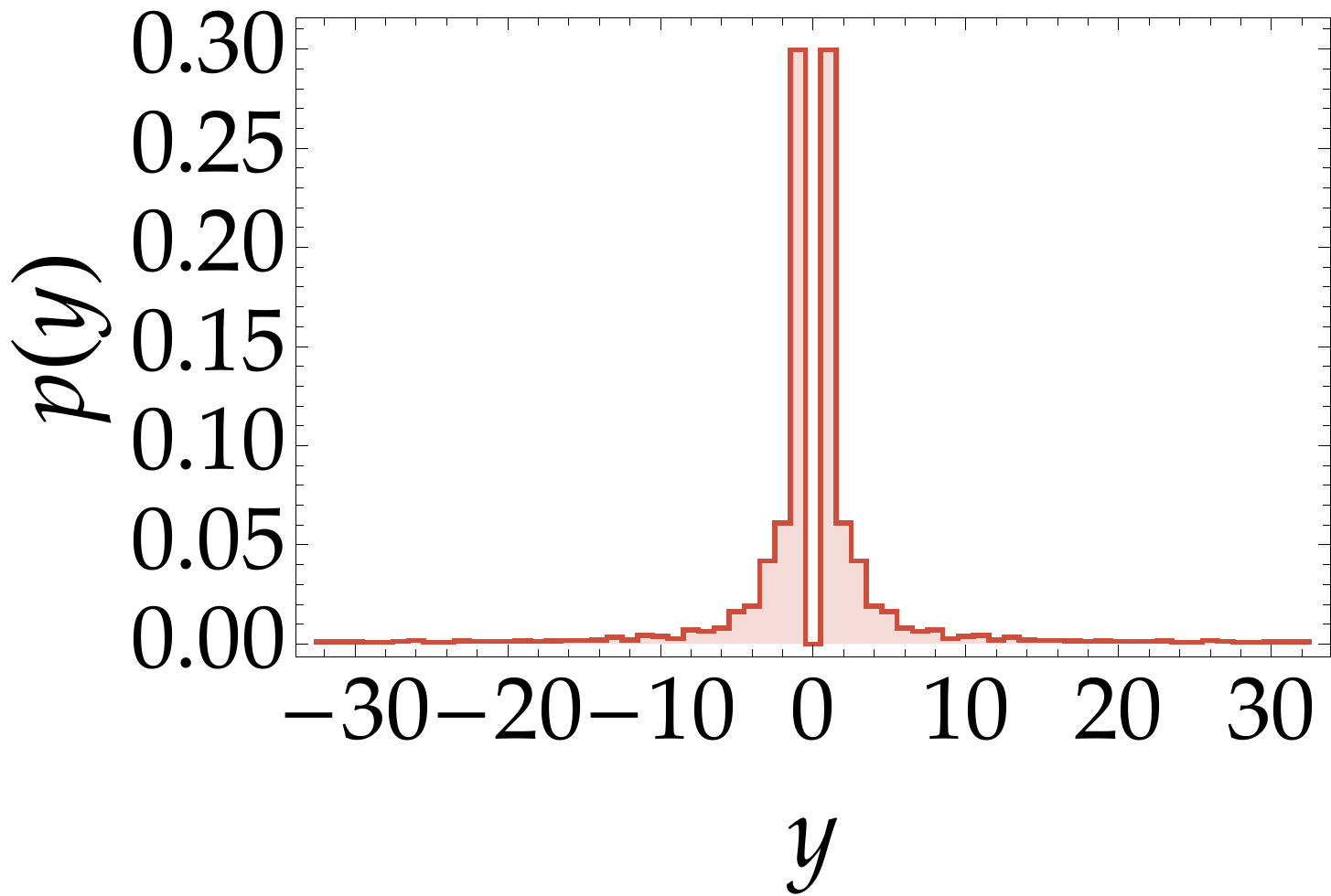}
  \caption{Probability ditribution in the relative coordinate $y$ of two proper
  eigenstates with vanishing total momentum and mass parameter $m=0.6$. On the
  left is shown the plot of the eigenstate $\int\!\dd k \, (\bv^{+-}_k -
  \bv^{-+}_k) e^{-iyk}$; on the right is shown the eigenstate $\int\!\dd k
  \, (\bv^{+-}_k + \bv^{-+}_k) e^{-iyk}$.}
  \label{fig:quit-bound}
\end{figure}


The solutions that we presented in the previous discussion do not cover the
extreme values $p=0,\, \pi/2$ (see Ref.~\cite{PhysRevA.97.032132} for a reference).
Let us consider for definiteness the case $p=0$ since the other case is obtained
in a similar way. For $e^{-i\omega} \neq 1$ the previous analysis still holds.
Indeed, noticing that $\omega_{\pm\pm}(0,k) = \pm 2 \omega(k)$, we have
$\omega(k) \in R$ and $\omega(k) \neq 0$ if and only if $k \in \Gamma_f \cup
\Gamma_0 \cup \Gamma_2$, whereas $\omega_{\pm\mp}(0,k) = 0$ for all $k \in \C$.
This means that the solutions $\ket{\psi_k^+}$ of \cref{eq:sol} are actually
eigenvectors of $U_2(\chi,0)$. Thus, the spectrum is made by a continuous part,
given by the arc of the unit circle containing $-1$ and having
$e^{\pm2i\omega(0)}$ as extremes, and a point spectrum with two points:
$e^{-2i\omega(\tilde k)}$, where $\tilde k$ is the solution of $T_+ = 0$ for
$p=0$, and $1$. As shown in Ref.~\cite{PhysRevA.97.032132}, $1$ is a separated part of
the spectrum of $U_2(\chi,0)$ and the corresponding eigenspace is a separable
Hilbert space of stationary bound states. This fact underlines an important
feature of the Thirring walk not shared by analogous Hamiltonian models. It is
remarkable that this behaviour occurs also for the free walk with $\chi=0$. In
\cref{fig:quit-bound} we show the probability distribution of two states having
he properties hereby discussed. It is worth noticing that all the states
$\bv^{+-}_k$ with $k\in(-\pi,\pi]$ are eigenvectors relative to the eigenvalue
$1$, and thus they generate a subspace on which the walk acts identically. We remark that this behaviour relies on the fact that the dispersion relation in one dimension is an even function of $k$.

In \cref{fig:bande} is depicted the discrete spectrum of the interacting walk
together with the continuous spectrum as a function of the total momentum $p$.
The solid curves in the gaps between the continuous bands denote the discrete
spectrum for different values of the coupling constant $\chi = 2\pi/3,\,
3\pi/7,\, -3\pi/7,\, -2\pi/3$. Molecule states appear also in the Hadamard walk
with the same on-site interaction~\cite{Ahlbrecht:2011ab}.

\section{Conclusions}

In this work we reviewed the Thirring quantum walk providing a simplified
derivation of its solutions for Fermionic particles. The simplified derivation
relies on the symmetric properties of the walk evolution operator allowing to
separate the subspace of solutions affected by the interaction from the subspace where the
interaction step acts trivially. The interaction term is the most general
number-preserving interaction in one dimension, whereas the free evolution is
provided by the Dirac QW~\cite{DAriano:2014ae}.

We showed the explicit derivation of the scattering solutions (solutions for the
continuous spectrum) as well as for the bound-state solutions. The Thirring walk
features also localized bound states (namely, states whose support is finite on
the lattice) when $e^{-i\omega} = e^{\pm i 2p}$. Such solutions exist only when
the coupling constant is $\chi = 2p$. In \cref{fig:loc-bound} is depicted the
evolution of a perfectly localized state showing the overlapping with bound
states components. In \cref{fig:bound-state} we reported the evolution of a
bound state of the two particles peaked around a certain value of the total
momentum: one can appreciate that the probability distribution remains localized
on the main diagonal during the evolution.

Finally, we discussed also the class of proper eigenvectors arising in the free
theory highlighting another difference between the discrete model of the present
work with analogous Hamiltonian models.

\vspace{6pt}


\acknowledgments{This publication was made possible through the support of a grant from the John Templeton Foundation under the project ID\# 60609 {\em Causal Quantum Structures}. The opinions expressed in this publication are those of the authors and do not necessarily reflect the views of the John Templeton Foundation.}

%

\appendixsections{multiple} 
\appendix

\section{Notation} \label{app:eigen-free}

For the single particle walk of \cref{eq:dirac-1d-op} the eigenstates can be
written as
\begin{equation*}
  \bv_p^s = \frac{1}{|N_s(p)|} \mvec{-i\mu \\ g_s(p)}, \qquad
    g_s(p) \deq -i(s\sin\omega(p) + \nu\sin p),
\end{equation*}
with $|N_s(p)|^2 = \mu^2 + |g_s(p)|^2$. For the two-particle walk we define
$\bv_k^{rs} \deq \bv_{p+k}^s \otimes \bv_{p-k}^r$. If $s=r$ than we name the
related eigenspace the \emph{even} eigenspace; whereas, if $s\neq r$ we call the
related eigenspace the \emph{odd} eigenspace. As proven in item $3$ of Lemma $1$
of Ref.~\cite{PhysRevA.97.032132}, for a given $k$ the degeneracy is $4$ both in the
even and in the odd case. Namely, if the triple $(+,+,k)$ is a solution then
also $(+,+,-k)$, $(-,-,\pi-k)$ and $(-,-,k-\pi)$ are solutions; if the triple
$(+,-,k)$ is a solution, then also $(+,-,\pi-k)$ and $(-,+,k-\pi)$ are
solutions.

Explicitly, for the even case we have:
\begin{align*}
  & \bv_k^{++} \propto
    \mvec{
      -\mu^2\\
      -i\mu g_+(p-k) \\
      -i\mu g_+(p+k) \\
      g_+(p+k) g_+(p-k)
    },
  & \bv_{\pi-k}^{--} \propto
    \mvec{
      -\mu^2\\
      i\mu g_+(p+k) \\
      i\mu g_+(p-k) \\
      g_+(p+k) g_+(p-k)
    }, \\[4ex]
  & \bv_{-k}^{++} \propto
    \mvec{
      -\mu^2\\
      -i\mu g_+(p-k) \\
      -i\mu g_+(p+k) \\
      g_+(p+k) g_+(p-k)
    },
  & \bv_{k-\pi}^{--} \propto
    \mvec{
      -\mu^2\\
      i\mu g_+(p-k) \\
      i\mu g_+(p+k) \\
      g_+(p+k) g_+(p-k)
    }.
\end{align*}
Analogously for the odd case the eigenstates are
\begin{align*}
  & \bv_k^{+-} \propto
    \mvec{
      -\mu^2\\
      -i\mu g_-(p-k) \\
      -i\mu g_+(p+k) \\
      g_+(p+k) g_-(p-k)
    },
  & \bv_{\pi-k}^{+-} \propto
    \mvec{
      -\mu^2\\
      i\mu g_+(p+k) \\
      i\mu g_-(p-k) \\
      g_+(p+k) g_-(p-k)
    }, \\[4ex]
  & \bv_{-k}^{-+} \propto
    \mvec{
      -\mu^2\\
      -i\mu g_+(p+k) \\
      -i\mu g_-(p-k) \\
      g_-(p+k) g_+(p-k)
    },
  & \bv_{k-\pi}^{-+} \propto
    \mvec{
      -\mu^2\\
      i\mu g_+(p+k) \\
      i\mu g_-(p-k) \\
      g_+(p+k) g_-(p-k)
    }.
\end{align*}
In order to simplify the derivation of the solution, we adopt the following
notation:
\begin{align*}
  & \bv_k^{++} \eqd \mvec{a\\b\\c\\d}, &
  & \bv_{-k}^{++} = \mvec{a\\c\\b\\d}, &
  & \bv_{\pi-k}^{--} = \mvec{a\\-c\\-b\\d}, &
  & \bv_{k-\pi}^{--} = \mvec{a\\-b\\-c\\d}, \\
  & \bv_k^{+-} \eqd \mvec{a'\\b'\\c'\\d'}, &
  & \bv_{-k}^{-+} = \mvec{a'\\c'\\b'\\d'}, &
  & \bv_{\pi-k}^{+-} = \mvec{a'\\-c'\\-b'\\d'}, &
  & \bv_{k-\pi}^{-+} = \mvec{a'\\-b'\\-c'\\d'}.
\end{align*}

\externalbibliography{yes}
\bibliography{bibliography}

\end{document}